\documentclass{elsart}
\usepackage{natbib}

\usepackage{epsf}

\newcommand{\lesssim}{\lower.5ex\hbox{$\; \buildrel < \over\sim \;$}}
\newcommand{\gtrsim}{\lower.5ex\hbox{$\; \buildrel > \over\sim \;$}}

\newcommand{\et}{\epsilon_T}

\begin{document}
\runauthor{Dermer and Atoyan}
\begin{frontmatter}
\title{Cosmic Rays and High-Energy Neutrinos from Gamma-Ray Bursts}
\author[NRL]{C.\ D.\ Dermer,}
\author[CRM]{A.\ Atoyan}

\address[NRL]{Code 7653, Naval Research Laboratory, 
Washington, DC 20375-5352 }
\address[CRM]{CRM, Universite de Montreal, Montreal H3C 3J7, Canada}

\begin{abstract}
Several lines of evidence point to a relationship between gamma-ray
bursts (GRBs) and the high mass stars that explode as
supernovae. Arguments that GRB sources accelerate cosmic rays (CRs)
are summarized. High-energy neutrino detection from individual GRBs
would mean that they are hadronically dominated, that is, that the
amount of energy deposited in the form of nonthermal hadrons in GRB
blast waves greatly exceeds the radiated energy inferred from their
photon emission spectra. Such a detection would make GRBs the favored
candidate sources of ultra-high energy and super-GZK CRs. Cascade
radiation induced by high-energy hadrons could be detected from GRBs with
$\gamma$-ray telescopes.
\end{abstract}
\end{frontmatter}

\section{Introduction}

Considerable evidence connects long-duration GRBs to star-forming
regions (see \cite{mes02,der02a} for review) and, consequently, to the
high-mass stars that radiate strongly in the UV band and evolve to
supernova core collapse.  For example, the host galaxies of GRBs have
blue colors, consistent with galaxy types that are undergoing active
star formation. GRB counterparts are found within the optical radii
and central regions of the host galaxies. Lack of optical
counterparts in approximately one-half of GRBs with well-localized
X-ray afterglows (so-called ``dark bursts") could be due to extreme
reddening from large quantities of gas and dust in star-forming host
galaxies. Supernova-like emissions have been detected in
the late-time optical decay curves of a few GRBs
including, most recently, the low redshift ($z =
0.17$) GRB 030329 \cite{sta03}.

Furthermore, X-ray features have been detected in about 6 GRBs, which
implies the existence of dense, highly enriched material in the
vicinity of the sources of GRBs (see \cite{bot02,laz02} for review).
This includes the variable absorption feature attributed to
photoionization of Fe in GRB 990705, K$\alpha$ and
recombination edges from highly ionized Fe and S in the afterglow
spectra of GRB 991216, multiple high-ionization emission features detected in
GRB 011211 and, at lower significance, claimed X-ray Fe
K$\alpha$ features in GRB 970508, GRB 970828, and GRB 000214.

The association of GRB 980425 and the Type Ic SN 1998bw \cite{pia00},
if true, directly connects GRBs and SNe.  The well-measured light
curve of SN 1998bw has been used in a number of cases to model
reddened excesses in the optical afterglow spectra found tens of days
after the GRB, again pointing to a SN/GRB connection.

Because supernovae are widely thought to accelerate galactic cosmic
rays, the supernovae associated with GRBs should play a special role
in the CR/GRB connection. Here we summarize work connecting CRs, SNe,
and GRBs, beginning with source models of GRBs and arguments for CR
acceleration by supernovae.  High-energy neutrino detection with
km-scale telescopes such as IceCube would conclusively establish that
hadrons are accelerated by GRBs, and is likely if GRB sources are
hadronically dominated by a factor of $\approx 10$ or more. This would
be required in a GRB model that accelerates nonthermal power-law
distributions of particles to produce ultra-high energy cosmic rays
(UHECRs) with energies $\gtrsim 10^{18.5}$ eV, and super-GZK CRs with
energies $\gtrsim 10^{20}$ eV. Hadronic cascade radiation could make
GRBs luminous sources of GeV-TeV emission that would be detectable
with {\it GLAST}, {\it VERITAS}, and {\it HESS}.

\section{Gamma-Ray Bursts and Supernovae}

Leading scenarios for the sources of long-duration GRBs are the
collapsar and supranova models.  In the former case \cite{woo93}, GRBs
are formed in the seconds to minutes following the collapse of the
core of a massive star to a black hole. During the collapse process, a
nuclear-density, several Solar-mass accretion disk forms and accretes
on the newly born black hole at the rate of $\sim 0.1$-$1~ M_\odot$
s$^{-1}$ to drive a baryon-dilute, relativistic outflow through the
surrounding stellar envelope. The duration of the accretion episode
corresponds to the period of activity of the relativistic winds that
are argued to produce the prompt variable GRB emission. A wide variety
of collapsar models are possible \cite{fwh99}, but their central
feature is the one-step collapse of the core of a massive star to a
black hole.  A major difficulty of this model is to drive a
baryon-dilute, relativistic outflow through the stellar envelope
\cite{tan01}.

In the supranova model \cite{vs98}, GRBs are the second in a two-step
collapse process of an evolved massive stellar core to a black hole
through the intermediate formation of a neutron star with mass
exceeding several Solar masses. The neutron star is initially
stabilized against collapse by rotation, but the loss of angular
momentum support through magnetic dipole and gravitational radiation
leads to collapse of the neutron star to a black hole after some weeks
to years. A two-step collapse process means that the neutron star is
surrounded by a SN shell of enriched material at distances of $\sim
10^{15}$-$10^{17}$ cm from the central source.  The earlier SN could
yield $\sim 0.1$-$1~M_\odot$ of Fe in the surrounding vicinity to
provide a surrounding shell of material that forms prompt and
afterglow X-ray features in GRB spectra.

A pulsar wind and pulsar wind bubble consisting of a quasi-uniform,
low density, highly magnetized pair-enriched medium within the SNR
shell is formed by a highly magnetized neutron star during the period
of activity preceding its collapse to a black hole \cite{kg02}. The
interaction of the pulsar wind with the shell material will fragment
and accelerate the SNR shell, and the pulsar wind emission will be a
source of ambient radiation that can be Comptonized to gamma-ray
energies \cite{igp02}. The heating of the SN shell by the plerionic
emission could give rise to a characteristic cooling shell signature,
which might explain some instances of the delayed reddened excesses in
optical afterglow spectra \cite{der02b}.

In both the collapsar and supranova models, the stellar progenitors
are $\gtrsim 10 $M$_\odot$ stars that evolve to core collapse and
produce a supernova with an expanding nonrelativistic shell, and a GRB
with its highly relativistic ejecta.  The shocks formed in these
nonrelativstic and relativistic outflows can accelerate cosmic rays.

\section{Cosmic Rays and Supernovae} 

The strongest argument that hadronic cosmic rays (CRs) are powered by
SNRs is probably the claim that only SNe inject sufficient power into
the Galaxy to provide the measured energy density of CRs
\cite{gs64,gai90}.  A time- and volume-averaged kinetic luminosity $\gtrsim
10^{41}$ ergs s$^{-1}$ in the disk of the Milky Way is required to
power the galactic CRs. The galactic SN luminosity is $ \approx (1$
SN$/30$ yrs$) \times 10^{51}$ ergs/SN $\approx 10^{42}$ ergs s$^{-1}$
which, given a 10\% efficiency for converting the directed kinetic
energy of SNe into CRs that seems feasible through the shock-Fermi
mechanism, is adequate to power the cosmic radiation.

Besides available power, Fermi acceleration at SNR shocks offers
another strong argument that cosmic rays originate from supernovae.
In Fermi models, the directed kinetic energy of a nonrelativistic or
relativistic outflow is transferred to a small fraction of
suprathermal particles, either through repeated scattering and
diffusion upstream and downstream of the shock front (first-order
Fermi acceleration), or through energy diffusion accompanying
stochastic pitch-angle scatterings with the magnetic turbulence
spectrum (second-order Fermi acceleration).

As compared with some other acceleration mechanisms, for example,
charge-depletion fields in pulsar magnetospheres or through magnetic
reconnection in stellar flares, the shock-Fermi mechanism naturally
produces power-law particle spectra, at least in the test particle
approximation. First-order Fermi acceleration at a nonrelativistic
strong shock gives the canonical test-particle shock index $s=-2$, and
first-order Fermi at a relativistic shock gives the canonical test
particle shock index $s \cong - 2.2$ \cite{ach01}. Modifying these
standard injection indices through energy-dependent diffusion for
impulsive, steady, or stochastic injection models offers a wide range of
flexibility to explain observations of cosmic rays.

\section{Cosmic Rays and Gamma-Ray Bursts}

Arguments for a connection between CRs and GRB sources are based on
the coincidence between available power of GRBs within the GZK radius
and the power required to accelerate super-GZK particles.  Moreover,
the available power from GRBs within our Galaxy is sufficient to
accelerate CRs between the knee at $\approx 3\times 10^{15}$ eV and
the ankle at $\approx 10^{18.5}$ eV (if these CRs are trapped in the
halo of our Galaxy), and could even power a significant fraction of the
GeV-TeV/nuc cosmic rays \cite{der02}.

The Larmor radius of a particle with energy $10^{20}E_{20}$ eV is
$\approx 100 E_{20}/(ZB_{\mu{\rm G}})$ kpc. Unless super-GZK CRs are
heavy nuclei, they probably originate from outside our Galaxy.  Based
on assumptions that the faintest BATSE GRBs are at redshifts $z
\approx 1$, it was shown \cite{vie95,wax95} that the energy density of
super-GZK CRs is comparable to the energy density that would be
produced by GRB sources within the GZK radius, assuming that the
measured $\gamma$-ray energy from GRBs is roughly equal to the total
energy deposited in the form of UHECRs.  This coincidence is verified
by a detailed estimate of GRB power in the context of the external
shock model for GRBs \cite{der02,bd00}. A prediction of this hypothesis is
that star-forming galaxies which host GRB activity will be surrounded
by neutron-decay halos.

The local density of $L^*$ galaxies like the Milky Way galaxy can be
derived from the Schechter luminosity function, and is $\approx
1/$(200-500 Mpc$^3$).  The BATSE observations imply $\sim 2$ GRBs/day
over the full sky. Due to beaming, this rate is increased by a factor
of 500 \cite{fra01}. This implies an event rate of about 1 GRB in the
Milky Way every $10^3$-$10^4$ years, and an energy injection rate of
$\approx 10^{40 \pm 1}$ ergs s$^{-1}$ \cite{der01}. This estimate takes into
account the assumption that the GRB rate density follows the
star-formation rate history of the universe, and a factor of $\approx
3$ to account for the contribution from clean and dirty fireballs,
such as the X-ray rich GRBs, for every classical long-duration GRB.
Thus the GRB rate is about 10\% as frequent as Type Ib/c SNe, and
about 1\% as frequent as Type II SNe.

Because each GRB has a total energy release of a few $\times 10^{51}$
ergs, a factor of a few greater than the energy release in ``normal"
SNe, but occurs $\sim 100$ times less frequently, the available power
from the sources of GRBs is a few per cent of the power from Type Ia
and Type II SNe.  The efficiency for accelerating CRs in the
relativistic outflows of GRBs could be much greater than in
nonrelativistic SNe, so that the progenitor sources of GRBs could make
a significant contribution to CRs at all energies.

From the rate estimates, we see that about 1 in 10 to 1 in 100 SNRs
would exhibit this enhanced emission from strong CR acceleration due
to an associated GRB.  The better imaging and sensitivity of the {\it
GLAST} telescope and the next generation of imaging ground-based air
Cherenkov telescopes, namely VERITAS, HESS and MAGIC, will test this
hypothesis.

\section{High-Energy Neutrinos from GRBs}

Detection of high-energy neutrinos from km-scale telescopes such as
IceCube will establish hadronic acceleration in sources such as
blazars and GRBs. We \cite{da03} have recently completed the first
quantitative set of calculations of photomeson neutrino production for
the collapsar and supranova models of GRBs that also treats associated
$\gamma\gamma$ pair-production attenuation of the emergent
$\gamma$-ray emission.  This model takes into account nonthermal proton
injection followed by photomeson energy loss, based on our
photo-hadronic model for blazar jets \cite{ad03}.  The basic
assumption in our calculations is that equal kinetic energy is
injected in nonthermal protons as is measured in hard X-ray and soft
$\gamma$-ray photons. Given the fluence of a GRB, the energy injected
in protons therefore depends only on the Doppler factor $\delta$.

Neutrinos will be formed through photomeson production with internal
synchrotron radiation and, in the supranova model, with external
synchrotron photons emitted by the pulsar wind electrons.
Even with the very strong radiation field that could occur in the
supranova model, model-independent arguments \cite{da03}, which we
reproduce here, show that it is not possible under the assumption of
equal energy in hadrons and observed radiation to detect neutrinos
from any GRBs except those exceedingly rare events at the level
$\gtrsim 3\times 10^{-4}$ ergs cm$^{-2}$,  which occur about
 2-5 times per year \cite{bri99}.

The detection efficiency in water or ice of ultrarelativistic
upward-going muon neutrinos ($\nu_\mu$) with energies $\epsilon_\nu =
10^{14}\epsilon_{T}$ eV is $P_{\nu\mu} \cong 10^{-4}\epsilon_T^\chi$,
where $\chi = 1$ for $\epsilon_T<1$, and $\chi =0.5$ for
$\epsilon_T>1$ \cite{ghs95}.  For a neutrino fluence spectrum
parameterized by $\nu \Phi_\nu =
10^{-4}\phi_{-4}\epsilon_T^{\alpha_\nu}$ erg cm$^{-2}$, the number of
$\nu_\mu$ detected with a km-scale $\nu$ detector such as IceCube with
area $A_\nu = 10^{10}A_{10}$ cm$^2$ is therefore
\begin{equation}
N_\nu(\geq \epsilon_T) \approx \int_{\et}^{\infty} {\rm d} \epsilon_1
  \;{\nu\Phi_\nu \over
\epsilon_{1}^{2}}\;P_{\nu\mu}A_\nu \simeq
0.6\;{\phi_{-4} A_{10}\over
{{1\over 2} - \alpha_\nu}}\cases{g_{\alpha_\nu}(\et) \;  ,& for
 $\epsilon_T < 1$ \cr\cr \et^{{\alpha_\nu} -1/2}\; , & for
$\epsilon_T \gtrsim 1$ ,\cr}\;\;
\label{Nnu}
\end{equation}
where $g_{\alpha_\nu}(\et) =1+[(1/ 2\alpha_\nu)-1](1-\et^{\alpha_\nu})$. For a $\nu\Phi_\nu$ spectrum with $\alpha_\nu \simeq 0$, the number of
$\nu_\mu$ to be expected are $N_\nu \simeq 1.2\phi_{-4} A_{10}
(1+{1\over 2} \ln \et^{-1})$ for $\et<1$, and $N_\nu \simeq
1.2\phi_{-4} A_{10} /\sqrt{\et}$ for $\et >1 $.  Thus, if the nonthermal
proton energy injected in the proper frame is comparable to the
radiated energy required to form GRBs with hard X-ray/soft
$\gamma$-ray fluences $\gtrsim 10^{-4}$ ergs cm$^{-2}$, then extremely
bright GRBs are required to leave any prospect for detecting $\nu_\mu$
with km-scale neutrino detectors. This estimate does not take into account the
loss of efficiency to produce $\nu_\mu$ from energetic protons through
photomeson processes, so the number of $\nu_\mu$ detected from
individual GRBs would be even less.

\begin{figure}[t]
\centerline{\epsfxsize=8cm  \epsfbox{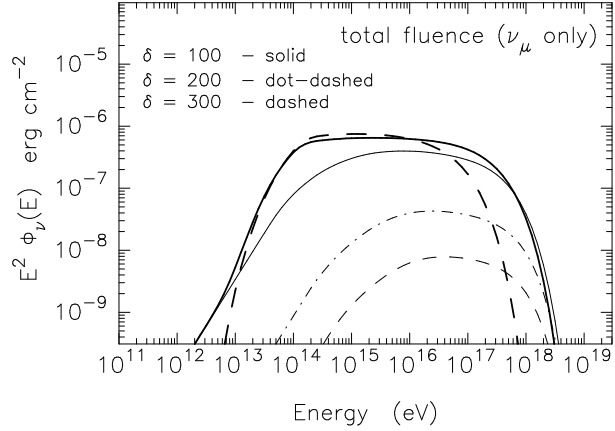}}
\caption{Energy fluence of photomeson muon neutrinos for a model 
GRB \cite{da03}. The thin
curves show collapsar model results where only the internal
synchrotron radiation field provides a source of target photons.  The
thick curves show the $\delta = 100$ and 300 results for the supranova
model calculation, which includes the effects of an external pulsar
wind radiation field.}
\end{figure}

Fig.\ 1 shows the total $\nu_\mu$ fluences expected from a model GRB
with a photon fluence of $3\times 10^{-5}$ ergs cm$^{-2}$ that is
radiated in 50 one-second pulses. The variability time scale determines
the size scale of the emitting region given $\delta$, and thus the
intensity of internal synchrotron photons for photomeson production.
The thin curves show collapsar model results at $\delta = 100,$ 200,
and 300. The expected numbers of $\nu_\mu$ that a km-scale neutrino
detector would detect are $N_\nu = 3.2\times 10^{-3}$, $1.5\times
10^{-4}$, and $1.9\times 10^{-5}$, respectively.  The heavy solid and
dashed curves in Fig.\ 1 give the supranova model predictions of
$N_\nu = 0.009$ for both $\delta = 100$ and $\delta = 300$. The
equipartition magnetic fields used in this calculation are 1.9 kG and
0.25 kG, respectively.  The external radiation field in the supranova
model makes the neutrino detection rate insensitive to the value of
$\delta$ (as well as to the variability time scale, as verified by
calculations). As the above estimate shows, there is no prospect to
detect $\nu_\mu$ from GRBs at this levels or from a GRB with a factor
of ten more fluence.

Although this calculation is not optimistic for neutrino detection,
note that the coincidence between the power released by GRBs in
$\gamma$ rays within the GZK radius and the power required for the
super-GZK CRs assumes that equal energy is injected into
the highest energy cosmic rays as is detected as GRB photon
radiation. This is possible in a second-order Fermi acceleration
scheme where a very hard, very high energy particle population is
produced \cite{dh01}. If particles are instead injected with a $-2$
or $-2.2$ spectrum through a first-order Fermi mechanism, then GRBs
must be hadronically dominated if they are to power the UHECRs. Thus,
$\approx 10$-100 times as much power must be injected in hadrons as
is inferred from the hard X-ray/soft $\gamma$-ray emission from
GRBs. Under these conditions, IceCube could expect to detect
high-energy neutrinos from several GRBs each year. Such a discovery
would have a profound impact on our understanding of cosmic rays. In
addition, luminous MeV-GeV-TeV cascade radiation induced by hadronic
secondaries would be produced. Observations of distinct $\gamma$-ray
components in GRB spectra
 would also support the GRB/CR connection.

\section{Conclusions}

Many lines of evidence point to a close linkage between supernovae,
gamma-ray bursts, and cosmic rays. Detection of high-energy neutrinos
from the sources of GRBs would confirm hadron acceleration to very
high energies. Under the assumption that equal energy is injected into
protons as is measured in radiation, we have found \cite{da03} that
there is no realistic prospect for neutrino detection except from the
most fluent GRBs at a level $\gtrsim 3\times 10^{-4}$ ergs cm$^{-2}$.
More optimistic estimates from the viewpoint of detecting GRB
neutrinos could be found in proton-dominated GRB models.  Without this
hypothesis only the brightest GRBs can be expected to be detected with
both high-energy $\gamma$-ray and neutrino detectors. The assumption
of hadronically dominated GRBs is required to provide the required
energy in UHECRs if the protons are accelerated in the form of a power
law with number index steeper than $-2$. Detection of high-energy
neutrinos from GRBs would test scenarios for GRBs, the hypothesis that
GRBs sources are powerful accelerators of ultrarelativistic CRs, and
our understanding of the origin of GRB radiation.

\vskip0.2in

The work of CDD is supported by the Office of Naval Research and NASA
{\it GLAST} science investigation grant DPR \# S-15634-Y.

\end{document}